\documentstyle[12pt]{article}
\begin{document}
\begin{titlepage}
\begin{center}
{\Large\bf Catalysis of proton decay in superstring theory\\}
\vspace*{15mm}
{\bf Yuri M. Malyuta\\}
\vspace*{10mm}
{\it Institute for Nuclear Research, National Academy of
Sciences of Ukraine\\
252022 Kiev, Ukraine\\}
{\bf  E-mail: aks@d310.icyb.kiev.ua}\\

\vspace*{40mm}
{\bf Abstract}
\end{center}
\large
\vspace*{1mm}
Theory of superstrings on Calabi-Yau manifolds is applied to
investigate catalysis of proton decay.
\end{titlepage}
\newpage
\large
\section{\bf   Introduction}
        Rubacov \cite{1.} and Callan \cite{2.} have shown
that grand unification theories
which do not conserve baryon number predict
the existence of mono\-po\-le--in\-duced
proton decay. Resently it has been proved that
Calabi-Yau models describe solitonic
objects of superstring theory  \cite{3.}.
The purpose of this work is to study induced
proton decay in the context of Calabi-Yau models.

During the last few years there has been substantial
progress in superstring
theory compactified on Calabi-Yau manifolds \cite{4.,5.}.
In particular the idea of
heterotic/type II string duality has proved
to be crucial. New methods enable to
investigate superstring vacua and phase transitions
between them. This approach
is very important because it gives new predictions
for future experimental searches.
\section{\bf  Catalysis reaction}
        Let us consider the instanton numbers (topological
invariants of rational
curves  \cite{4.}) for the Calabi-Yau manifolds
$  X_{24}(1,1,2,8,12)^{3,243}_{-480} $ and
$  X_{20}(1,1,2,6,10)^{4,190}_{-372}  $ recorded in Tables 1 and 2.
\vspace*{7mm}
\begin{center}
{\bf Table 1}

\vspace*{3mm}
\begin{tabular}{|lllccrrr|}         \hline
$ (0,0,1) $  & $ \hspace*{4mm}    -1    $  & $  (0,1,1) $  & $
\hspace*{4mm}    -1  $  &
$ (0,1,2) $ & $   -2 $  & $ (0,1,3) $ & $  -3  $ \\
$ (0,1,4) $  & $ \hspace*{4mm}    -4    $  & $  (0,1,5) $  &
$ \hspace*{4mm}    -5  $  & $ (0,2,3) $ & $   -3 $  &
$ (0,2,4) $ & $ -16  $ \\
$ (1,0,0) $  & $ \hspace*{2.5mm}   240  $  & $  (1,0,1) $  &
$ \hspace*{2mm}   240  $  & $ (1,1,1) $ & $  240 $  & $ (1,1,2)
$ & $ 720  $ \\
$ (1,1,3) $  & $                1200    $  & $  (1,1,4) $  &
$   1680               $  & $ (1,2,3) $ & $ 1200 $  & $ (2,0,0) $ &
$ 240  $ \\
$ (2,0,2) $  & $ \hspace*{2.5mm}   240  $  & $  (2,2,2) $  &
$ \hspace*{2mm}   240  $  & $ (3,0,0) $ & $  240 $  & $ (3,0,3) $ &
$ 240  $ \\
$ (4,0,0) $  & $ \hspace*{2.5mm}   240  $  & $  (5,0,0) $  &
$ \hspace*{2mm}   240  $  & $ (6,0,0) $ & $  240 $  &
$ (0,1,0) $ & $   0  $ \\   \hline
\end{tabular}\\
\vspace*{4mm}
{\bf Table 2}

\vspace*{3mm}
\begin{tabular}{|lllccrrr|}         \hline
$(0,0,0,1) $ & $ \hspace*{4.5mm}  28  $ & $ (0,0,0,2) $ &
$ \hspace*{6mm}    -1  $ & $ (0,0,0,3) $ & $  0 $ &
$ (0,0,1,0) $ & $  -1 $\\
$(0,1,0,0) $ & $ \hspace*{7mm}   0  $ & $ (0,1,1,0) $ &
$ \hspace*{6mm}    -1  $ & $ (0,1,2,0) $ & $ -2 $ &
$ (0,1,3,0) $ & $  -3 $\\
$(0,1,4,0) $ & $ \hspace{3mm}   -4  $ & $ (0,1,5,0) $ &
$ \hspace*{6mm}    -5  $ & $ (0,2,3,0) $ & $ -3 $ &
$ (0,2,4,0) $ & $ -16 $\\
$(0,2,5,0) $ & $  -55           $     & $ (0,2,6,0) $ &
$  -144                $ & $ (0,3,4,0) $ & $ -4 $ &
$ (0,3,4,0) $ & $  -4 $\\
$(0,3,5,0) $ & $  -55               $ & $ (1,0,0,0) $ &
$ \hspace*{6mm}    -1  $ & $ (1,0,0,1) $ & $ 28 $ &
$ (1,0,0,2) $ & $ 186 $\\
$(1,0,0,3) $ & $ \hspace*{4mm}  28  $ & $ (1,0,0,4) $ &
$ \hspace*{6mm}    -1  $ &             &      &             & \\  \hline
\end{tabular}
\end{center}
\vspace*{4mm}
The calculation of these numbers was done using
the computer program INSTANTON.
>From Tables 1 and 2 we infer the following relation
between the instanton numbers
\begin{equation}
  n_{a,b,c}=\sum_{k}n_{a,b,c,k}.
\end{equation}
Relation (1) describes the phase transition between
solitonic objects with the
spectra presented in Table 3.
\begin{center}
{\bf Table 3}
\end{center}
\begin{center}
\begin{tabular}{|l|l|r|r|}  \hline
\hspace*{8mm} Calabi-Yau                 &
\hspace*{3mm}    spectrum                                  &
gauge group \hspace*{1mm} &  rank of group  \\ \hline
                                         &                                                           &                              &                 \\
$   X_{24}(1,1,2,8,12)^{3,243}_{-480} $  &
\hfill $  244 \hspace*{3mm} {\bf 1}                      $ &
$  U(1)^{4} $                &    $  4 $ \hspace*{11mm}      \\
                                         &                                                           &                              &                 \\
$   X_{20}(1,1,2,6,10)^{4,190}_{-372} $  &
$   28 \hspace*{3mm}{\bf 2} + 191 \hspace*{3mm} {\bf 1} $ &
$ SU(2)\times{U(1)^{4}} $    &    $  5 $ \hspace*{11mm}      \\
                                         &                                                           &                              &                 \\  \hline
\end{tabular}
\end{center}
\vspace*{5mm}

These spectra are calculated by using the technique of
instanton numbers  \cite{5.}:\\

\hspace*{33mm}      $            240 \hspace*{3mm}   =  -1 +
\framebox[10mm]{28}   + 186 +  \framebox[10mm]{28}  - 1   $  \\
 \hspace*{44mm}     4  +1 \hspace*{32mm}             5               \\
 \hspace*{37mm}  $  \overline  {\framebox[10mm]{244}+1}
\hspace*{28mm} \overline   {\framebox[10mm]{191}}  $  \hfill (2)\\
\vspace*{5mm}\\
We conclude from (2) that the solitonic object
$  Sol = (244\hspace{3mm} {\bf 1})  $  consists
of  244 singlet hypermultiplets, while the solitonic
object  $  Sol^{*} = (28 \hspace*{3mm}{\bf 2} + 191 \hspace*{3mm}{\bf 1}) $
consists of 28 doublet hypermultiplets and 191 singlet hypermultiplets;
moreover, there exists the phase transition\\
\vspace*{2mm}\\
\hspace*{50mm} $  Sol^{*} \rightarrow Sol + H ,$  \hfill    (3)\\
\vspace*{2mm}\\
where $ H $ is the singlet hypermultiplet.

        Combining (3) with the quark diagram of proton
decay we construct the diagram of the catalysis
\vspace*{2mm}\\
\hspace*{63mm}  $  \vspace*{-5.4mm} \longrightarrow   \hspace*{1mm}   $ \\
\hspace*{59mm}  $  p \vspace*{-5.4mm} \longrightarrow
\hspace*{1mm}  \bigcirc  \hspace*{1mm}  \longrightarrow  e^{+}   $   \\
\hspace*{63mm}  $  \vspace*{-5.4mm} \longrightarrow  \hspace*{1mm}  $  \\
\vspace*{-1mm}\\
\hspace*{74mm}  $   \downarrow \hspace*{-1.2mm}  \uparrow  $
{\small $ H $ }  \\
\vspace*{-5.5mm}\\
\hspace*{52mm}  $  Sol^{*} \Longrightarrow
\hspace*{1mm} \bigcirc  \hspace*{1mm} \Longrightarrow Sol  $ \hfill (4) \\
\section{\bf Estimates of bounds}
Reaction (4) has the energy emission $  \varepsilon  \simeq  m_{H} $
where we equate $  m_{H}  $  with  the Higgs mass
$ 1000 $ GeV \cite{6.}. To estimate
the cross section of reaction (4) let us apply the
Okun trick \cite{7.}. This trick is
demonstrated in  Table 4  where $  \alpha_{12} = g_{1}g_{2} $ , $ g_{1} $
and $  g_{2} $ are effective coupling constants
which represent the vertices
of the diagram (4).\\
\begin{center}
{\bf Table 4}\\
\end{center}
\begin{center}
\begin{tabular}{|l|c|r|}  \hline
\hspace*{1mm}Amplitude                       &
Probability \hspace*{3mm}                            &
Cross section \hspace*{6mm}         \\
of reaction (4)                              &
of reaction (4)                                &
of reaction (4) \hspace{4mm}              \\  \hline
                                             &                                                      &                                                          \\
  $ A=\alpha_{12}m^{-2}_{H} $                &
$ \omega=A^{2}\varepsilon^{5}=\alpha^{2}_{12}m_{H} $ &
$ \sigma=\omega^{-2}=(\alpha^{2}_{12}m_{H})^{-2} $ \\
                                             &                                                      &                                                   \\    \hline
\end{tabular}
\end{center}
\vspace*{5mm}

        Identifying in the formula for  $ \sigma $
the constant   $ \alpha_{12} $  with the grand unification
constant   $ \alpha_{GU}\simeq\frac{1}{40}  $ \cite{7.}
we obtain the upper bound of the catalysis cross section
$  \sigma\simeq10^{-28}cm^{2} $ .

\end{document}